\documentclass[12pt]{article}

\usepackage{graphicx}
\usepackage{amssymb,amsmath}
\usepackage{subfigure}
\usepackage{color}

\graphicspath{{./figures/}}

\makeatletter

\def\ve#1{{\mathchoice{\mbox{\boldmath$\displaystyle #1$}}%
              {\mbox{\boldmath$\textstyle #1$}}%
              {\mbox{\boldmath$\scriptstyle #1$}}%
              {\mbox{\boldmath$\scriptscriptstyle #1$}}}}
\newcommand{\E}{{\mathrm{\sf E}}}
\newcommand{\e}{{\mathrm{\sf e}}}
\newcommand{\ber}{{\mathrm{BER}}}
\newcommand{\Es}{{\bar{E}_s}}
\renewcommand{\H}{H^{\mathrm{n}}}
\newcommand{\vH}{\ve{H}^{\mathrm{n}}}
\renewcommand{\j}{{\mathrm{\sf j}}}

\def\section{\@startsection{section}{1}{\z@}
     {-1.5ex plus-1ex minus -.2ex}{2.0ex minus 2.0ex}
     {\Large\sf\textbf}}
\def\subsection{\@startsection{subsection}{2}{\z@}
     {-2.5ex plus-1ex minus -.2ex}{1.5ex plus.2ex}
     {\large\sf\textbf}}
\def\subsubsection{\@startsection{subsubsection}{3}{\z@}
     {-1.25ex plus-1ex minus -.2ex}{1.5ex plus.2ex}
     {\normalsize\sf\textbf}}
\def\paragraph{\@startsection{paragraph}{4}{\z@}
     {-1.25ex plus-1ex minus -.2ex}{1.5ex plus.2ex}
     {\normalsize\sf\textsl}}

\newcommand{\Section}[1]{\vspace{0mm}\section{#1}\vspace{0mm}} 
\newcommand{\SubSection}[1]{\vspace{0mm}\subsection{#1}\vspace{0mm}} 
\newcommand{\SubSubSection}[1]{\vspace{0mm}\subsubsection{#1}\vspace{0mm}}

\def\ps@headings{\let\@mkboth=\markboth
        \def\@oddhead{\vbox{\hsize\textwidth \hbox to \textwidth{%
                        \small\sf %\today
                        \textsl{Snow et al.: Performance Analysis and Enhancement of Multiband OFDM for UWB Communications}%
                        \hfil \thepage}
                        \vskip 3pt \hrule}\hss%
        }
        \def\@oddfoot{\hss%
        }
}
\makeatother

\long\def\symbolfootnote[#1]#2{\begingroup%
\def\thefootnote{\fnsymbol{footnote}}\footnote[#1]{#2}\endgroup} 

\begin{document}
\sf

\thispagestyle{empty}
\vspace*{-20mm}
\begin{center}
{\Huge Performance Analysis and Enhancement of \\[0.3em] Multiband OFDM for UWB Communications$^*$}
\end{center}
\vspace*{5mm}

\begin{center}
Chris Snow$^\dagger$, Lutz Lampe, and Robert Schober \\[0.5em]
Department of Electrical and Computer Engineering \\
The University of British Columbia \\
Vancouver, British Columbia, Canada \\
Email: $\{$csnow,lampe,rschober$\}$@ece.ubc.ca
\end{center}
\vspace*{6mm}

\renewcommand{\baselinestretch}{1.5}
\large\normalsize

{{\slshape Abstract}} --- In this paper, we analyze the frequency-hopping orthogonal frequency-division multiplexing (OFDM) system known as \textsl{Multiband OFDM} for high-rate wireless personal area networks (WPANs) based on ultra-wideband (UWB) transmission. Besides considering the standard, we also propose and study system performance enhancements through the application of Turbo and Repeat-Accumulate (RA) codes, as well as OFDM bit-loading. Our methodology consists of (a) a study of the channel model developed under IEEE 802.15 for UWB from a frequency-domain perspective suited for OFDM transmission, (b) development and quantification of appropriate information-theoretic performance measures, (c) comparison of these measures with simulation results for the Multiband OFDM standard proposal as well as our proposed extensions, and (d) the consideration of the influence of practical, imperfect channel estimation on the performance. We find that the current Multiband OFDM standard sufficiently exploits the frequency selectivity of the UWB channel, and that the system performs in the vicinity of the channel cutoff rate. Turbo codes and a reduced-complexity clustered bit-loading algorithm improve the system power efficiency by over 6~dB at a data rate of 480~Mbps.

\vspace*{5mm}

{\slshape Index terms:} Ultra wideband, Multiband OFDM, diversity, bit loading, information-theoretic measures

\vspace*{8mm}

$^*$ {\small This work was presented in part at the IEEE International Conference on Communications (ICC), Seoul, May 2005, and in part at the IEEE International Conference on Ultra-Wideband (ICU), Zurich, September 2005. This work has been supported in part by the National Science and Engineering Research Council of Canada (Grant CRDPJ 320552) and Bell University Laboratories, and in part by a Canada Graduate Scholarship.}

$^\dagger$ {\small Corresponding author}

%%%%%%%%%%%%%%%%%%%%%%%%%%%%%%%%%%%%%%%%%%%%%%%%%%%%%%%%%%%%%%%%%%%%%%%%%%%%%%%
\newpage
\renewcommand{\baselinestretch}{1.58}\normalsize
\setcounter{page}{1}

\Section{Introduction}
\vspace{-5mm}

Ultra-wideband (UWB) radio has recently been popularized as a technology for short-range, high data rate communication and locationing applications (cf.\ e.g.\ \cite{RFSL04}). The IEEE 802.15 standardization group, responsible for wireless personal area networks (WPANs), organized task group 3a to develop an alternative physical layer based on UWB signaling \cite{WPAN3a}. There were two main contenders for this standard: a frequency-hopping orthogonal frequency-division multiplexing (OFDM) proposal known as \textsl{Multiband OFDM} and a code-division multiple access (CDMA) based technique. 

In this paper, we consider the proposed Multiband OFDM standard \cite{MBOFDM} (also recently standardized by the ECMA~\cite{ECMA}). Multiband OFDM is a conventional OFDM system \cite{bingham:2000} combined with bit-interleaved coded modulation (BICM) \cite{CaireEtAl98} for error prevention and frequency hopping for multiple access and improved diversity. The signal bandwidth is 528~MHz, which makes it a UWB signal according to the definition of the US Federal Communications Commission (FCC) \cite{RFSL04}, and hopping between three adjacent frequency bands is employed for first generation devices \cite{MBOFDM}. Thus, the Multiband OFDM proposal is a rather pragmatic approach for UWB transmission, which builds upon the proven BICM-OFDM concept.\footnote{\sf Throughout this paper, the term ``Multiband OFDM'' refers to the particular standard proposal \cite{MBOFDM}, whereas ``BICM-OFDM'' refers to the general concept of combining BICM and OFDM.}

The objective of this paper is to study the suitability and to analyze the (potential) performance of Multiband OFDM for UWB transmission. Furthermore, we propose system performance enhancements by applying capacity-approaching Turbo and Repeat-Accumulate (RA) codes and by using OFDM bit-loading. These specific techniques were chosen because of their potential for improved system performance without requiring substantial changes to other portions of the Multiband OFDM system, nor requiring major increases in complexity. Since our investigations rely on the new UWB channel model developed under IEEE 802.15 \cite{MolischEtAl03}, we first analyze this channel model in the frequency domain and extract the relevant statistical parameters that affect the performance of OFDM based transmission. In particular, the amount of diversity available in the wireless channel as a function of the signal bandwidth is examined. As appropriate performance measures for coded communication systems, we discuss the capacity and cutoff rate limits of BICM-OFDM systems for UWB channels. In this context, since one limiting factor of performance in practical and especially in wideband BICM-OFDM systems is the availability of high-quality channel state estimates, the effect of imperfect channel state information (CSI) at the receiver is specifically addressed. Furthermore, the information-theoretic performance limits are compared with simulated bit-error rate (BER) results for the Multiband OFDM proposal and the extensions introduced herein.

The literature on Multiband OFDM systems and their performance is relatively sparse. In \cite{BatraEtAl04} the authors present an overview of the Multiband OFDM system as well as performance results. The authors of \cite{Wessman04} consider properties of the UWB channel related to Multiband OFDM, and independently arrive at results agreeing with our findings in Section \ref{sec:marginaldist}.\footnote{\sf We note that our conference paper (presented at ICC 2005) was submitted before \cite{Wessman04} appeared.} As an extension to the standard proposal, simplified Low-Density Parity-Check (LDPC) codes are considered in \cite{PngEtAl04} in order to improve the power efficiency of the Multiband OFDM system for a subset of the proposed data rates. The authors of \cite{XuLiu04} consider the application of a clustered power allocation scheme to Multiband OFDM. However, this scheme attempts to maximize throughput and therefore does not provide fixed data rates compatible with the Multiband OFDM standard proposal. In \cite{uwb:SSOL05} the authors present a space-time-frequency coding scheme for Multiband OFDM. A subband and power allocation strategy for a multiuser Multiband OFDM system is given in \cite{uwb:SHL05}, but each user in the system uses a fixed modulation (i.e. no per-user bit allocation is performed). We note that none of these previous works %\cite{BatraEtAl04,Wessman04,PngEtAl04,XuLiu04,uwb:SSOL05,uwb:SHL05} 
provide comparisons with relevant information-theoretic limits.

The remainder of this paper is organized as follows. Section~\ref{sec:models} describes the Multiband OFDM system and the performance enhancements we propose, as well as the UWB channel model under consideration. The properties of the UWB channel relevant for OFDM transmission are examined in Section \ref{sec:Channel}. Section~\ref{sec:capeval} presents the capacity and cutoff rate analysis and numerical results. Simulation results for the Multiband OFDM system and the proposed extensions are presented and compared with the theoretical benchmark measures in Section~\ref{sec:simresults}, and conclusions are given in Section~\ref{sec:conclusions}.

%%%%%%%%%%%%%%%%%%%%%%%%%%%%%%%%%%%%%%%%%%%%%%%%%%%%%%%%%%%%%%%%%%%%%%%%%%%%%%%
\Section{Multiband OFDM, Extensions, and UWB Channel Model}
\label{sec:models}

In this section, the transmission system and channel model are introduced. We describe the transmitter of the proposed Multiband OFDM standard \cite{MBOFDM} as well as extensions to channel coding and to modulation. For the receiver we adopt a conventional state-of-the art architecture including channel estimation based on pilot symbols.

\SubSection{Transmitter}
\label{sec:mbofdm}
\vspace{-3mm}

\begin{figure}[tbp]
\centering
        \subfigure[\sf{Transmitter}]{
        \label{fig:sysdiag_transmitter}
        \includegraphics[width=0.85\columnwidth]{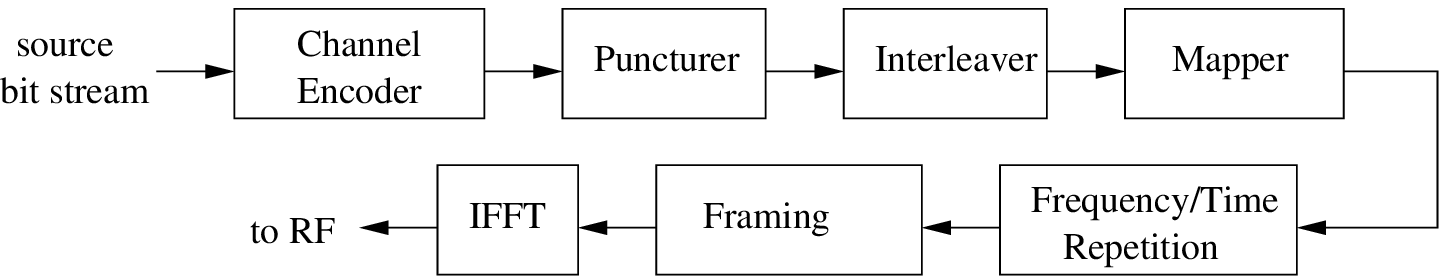}}
        \subfigure[\sf{Receiver}]{
        \label{fig:sysdiag_receiver}
        \includegraphics[width=1\columnwidth]{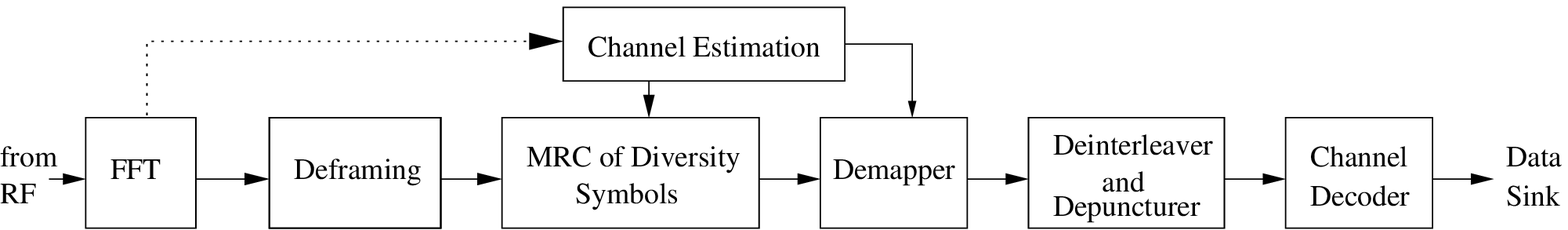}}
        \caption{\sf\label{fig:sysdiag} %
        Block diagram of Multiband OFDM transmission system.}
\end{figure}

The block diagram of the Multiband OFDM transmitter is shown in Fig~\ref{fig:sysdiag}a). A total of ten data rates (from 53.3 Mbps to 480 Mbps) are supported by the use of different code puncturing patterns as well as time and/or frequency repetition. We present a description of the Multiband OFDM standard proposal \cite{MBOFDM} in parallel with our extensions to channel coding and to modulation.

\SubSubSection{Channel Coding and Spreading}
\vspace{-2mm}

\textsl{\indent A. Multiband OFDM Standard Proposal:} Channel coding in the proposed standard consists of classical BICM \cite{CaireEtAl98} with a punctured maximum free distance rate $R=1/3$ constraint length 7 convolutional encoder. A multi-stage block-based channel interleaver is used (see \cite{MBOFDM} for details). After modulation (described below), modulated symbols are optionally repeated in time (in two consecutive OFDM symbols) and/or frequency (two tones within the same OFDM symbol), reducing the effective code rate by a factor of 2 or 4 and providing an additional spreading gain for low data rate modes. The channel interleaver length (300, 600 or 1200 coded bits) depends on the spreading factor.

\textsl{B. Extension --- Turbo Codes:} We propose the use of Turbo codes \cite{Berrou96} in order to improve the system power efficiency and more closely approach the channel capacity. We examined generator polynomials of constraint length 3, 4 and 5 as well as various interleavers (including s-rand \cite{DolinarDivsalar95} and dithered relative prime \cite{CrozierGuinand01} designs). Due to their excellent performance for the code lengths considered as well as reasonable interleaver memory storage requirements, we decided to adopt the generator polynomials and interleaver design developed by the 3rd Generation Partnership Project (3GPP) \cite{3gpp}. For low data rates, the time/frequency spreading technique of the Multiband OFDM proposal is retained. We would like to maintain compatibility with the Multiband OFDM channel interleaver by having each coded block fit into one channel interleaver frame, as is done with the convolutional codes used in the standard.\footnote{\sf Note that keeping the block lengths short also reduces the memory requirements and decoding delay at the receiver.} However, to maintain compatibility at the lowest data rates would require a Turbo code interleaver length of only 150 or 300 bits. Due to the poor distance properties and resultant performance degradation associated with short-length Turbo codes, at low data rates we consider both Multiband OFDM-compliant block lengths and longer blocks of 600 input bits (the same length as used without spreading).

\textsl{C. Extension --- RA Codes:} The limited length of the Multiband OFDM channel interleaver motivates the consideration of serially-concatenated codes, where the interleaver is positioned between the constituent encoders and thus has a longer length. We consider nonsystematic regular RA codes \cite{Divsalar98} due to their simplicity and good performance for the required code lengths. The time/frequency spreading mechanism described above is discarded, and low-rate RA codes ($R=1/4$ or $1/8$) are used. The interleaver between the repeater and accumulator is randomly generated (no attempt is made to optimize its performance).

\SubSubSection{Modulation}
\label{sec:modulation}
\vspace{-2mm}

\textsl{\indent A. Multiband OFDM Standard Proposal:} In the proposed standard, the interleaved coded bits are mapped to quaternary % this is the correct spelling
phase-shift keying (QPSK) symbols using Gray labeling. After the optional spreading described above, groups of 100 data symbols are used to form OFDM symbols with $N=128$ tones.

\textsl{B. Extension --- Bit-Loading:} The UWB channel (see Section \ref{sec:chanmodel}) is considered time-invariant for the duration of many packet transmissions. For that reason, it is feasible to consider bit-loading algorithms to assign unequal numbers of bits to each OFDM subcarrier \cite{bingham:2000}. Channel state information is obtained at the transmitter, either by (a) exploiting channel reciprocity (if the same frequency band is used in the uplink and downlink as in the standard proposal), or (b) some form of feedback (which may be required even if the same frequency band is used, since reciprocity may not apply due to different interference scenarios for transmitter and receiver). We consider loading for higher data rates (without time or frequency spreading) using two different OFDM bit-loading schemes. We selected the algorithm of Piazzo \cite{Piazzo} (which loads according to the uncoded BER) due to its low computational complexity, and the algorithm of Chow, Cioffi and Bingham (CCB) \cite{CCB} because it loads according to the information-theoretic capacity criterion, as well as for its moderate computational complexity. 

The data rates and OFDM symbol structure of the Multiband OFDM proposal are maintained by loading each OFDM symbol with 200 bits. Each tone carries from 0 to 6 bits using Quadrature Amplitude Modulation (QAM) signal constellations with Gray or quasi-Gray labeling (note that 6 bit/symbol corresponds to 64-QAM, which is a reasonable upper limit for modulation on a wireless channel). Due to FCC restrictions on the transmitted power spectral density, power loading is not used (all tones carry the same power). The target uncoded BER for the Piazzo algorithm is chosen $10^{-5}$ (cf. \cite{Piazzo} for details), but we found that performance is quite insensitive to this parameter. For the CCB algorithm, the signal-to-noise ratio (SNR) gap parameter $\Gamma$ is either 6~dB (when convolutional codes are used) or 3~dB (for Turbo codes). When the algorithm is unable to determine a suitable loading an all-QPSK loading is used, cf. \cite{CCB} for details.

\textsl{C. Extension --- Clustered Bit-Loading:} One potential feedback-based method of bit-loading is for the receiver to determine the appropriate modulation for each tone and feed the loading information back to the transmitter. To lower the feedback transmission requirements and significantly reduce the loading algorithm's computational complexity, we propose a clustered loading scheme where clusters are formed by considering groups of $D$ adjacent tones. As we found the CCB algorithm superior over the Piazzo algorithm in terms of achievable power efficiency (see Sections~\ref{sec:noclustercapcutres} and~\ref{sec:sim_loading}), we make the following modification to the CCB algorithm. We substitute Eq. (1) of \cite{CCB} with:
\begin{equation}
b(i) = \frac{1}{D}\sum_{k=1}^{D}\log_2\left(1 + SNR(i,k)\cdot10^{-\left(\frac{\Gamma + \gamma_{margin}}{10}\right)}\right)
\end{equation}
where $SNR(i,k)$ is the signal-to-noise ratio of the $k^{\mathrm{th}}$ tone in the $i^{\mathrm{th}}$ cluster, $\gamma_{margin}$ is the system performance margin in dB (iteratively calculated by the CCB algorithm), and $b(i)$ is the (possibly non-integer) number of bits allocated for each tone in cluster $i$. Using the modified algorithm to load $200/D$ bits on $100/D$ clusters provides the final integer-valued loadings $\hat{b}(i)$ for each cluster. Finally, all tones in cluster $i$ are assigned $\hat{b}(i)$ bits (i.e. the loading inside each cluster is constant). This modification causes the CCB algorithm to load according to the mean capacity of all tones in a cluster.

\SubSubSection{Framing and Transmission}
\vspace{-2mm}

The time domain signal is generated via inverse fast Fourier transform (IFFT) and a cyclic prefix of 32 samples is appended. The radio frequency (RF) transmit signal hops after each OFDM symbol between three 528~MHz frequency bands with center frequencies at 3.432, 3.960, and 4.448~GHz (see \cite{MBOFDM} for more details). Transmission is organized in packets of varying payload lengths. Each packet header contains two pilot OFDM symbols (all tones are pilots) per frequency band, which are used at the receiver to perform channel estimation (see Section \ref{sec:chanestalgo}).

\SubSection{UWB Channel Model}
\label{sec:chanmodel}
\vspace{-3mm}

For a meaningful performance analysis of the Multiband OFDM proposal, we consider the channel model developed under IEEE 802.15 for UWB systems \cite{MolischEtAl03}. The channel impulse response is a Saleh-Valenzuela model~\cite{Saleh+Valenzuela87} modified to fit the properties of measured UWB channels. Multipath rays arrive in clusters with exponentially distributed cluster and ray interarrival times. Both clusters and rays have decay factors chosen to meet a given power decay profile. The ray amplitudes are modeled as lognormal random variables, and each cluster of rays also undergoes a lognormal fading. To provide a fair system comparison, the total multipath energy is normalized to unity. Finally, the entire impulse response undergoes an ``outer'' lognormal shadowing.  The channel impulse response is assumed time invariant during the transmission period of several packets (see \cite{MolischEtAl03} for a detailed description).

Four separate channel models (CM1-CM4) are available for UWB system modeling, each with arrival rates and decay factors chosen to match a different usage scenario. The four models are tuned to fit 0-4~m Line-of-Sight (LOS), 0-4~m non-LOS, 4-10~m non-LOS, and an ``extreme non-LOS multipath channel'', respectively. The means and standard deviations of the outer lognormal shadowing are the same for all four models. The model parameters can be found in~\cite[Table 2]{MolischEtAl03}.

\vspace{-2mm}
\SubSection{Receiver}
\label{sec:mbofdmrec}
\vspace{-3mm}

The block diagram of the receiver considered in this paper is depicted in Fig~\ref{fig:sysdiag}b). We assume perfect timing and frequency synchronization. For this paper (with the exception of Section~\ref{sec:marginaldist}), we consider only channels CM1-CM3, where the cyclic prefix is longer than the delay spread of the channel impulse response.\footnote{\sf However, we note that the CM4 performance is very similar to that of CM3.} Thus, after FFT we see an equivalent $N$ dimensional frequency non-selective vector channel, expressed as \cite{bingham:2000},
\begin{equation}
\label{eq:ofdmsys}
\ve{Y}[k] =\ve{X}_d[k]  \ve{H} + \ve{N}[k]\;,
\end{equation}
where the vector notation $\ve{Z}[k] = [Z_1[k]\ldots Z_{N}[k]]^T$ is used ($\cdot^T$ denotes transpose) and $\ve{X}_d[k]$ is the $N\times N$ diagonal matrix with elements $X_i[k]$ at its main diagonal. $Y_i[k]$, $X_i[k]$, and $N_i[k]$ are the received symbol, the transmitted symbol,  and the additive white Gaussian noise (AWGN) sample on frequency tone $i=1\ldots N$ of the $k$th OFDM symbol, respectively. The vector $\ve{H}$ contains the frequency domain samples of the channel transfer function on tones $i=1\ldots N$ and is assumed constant over the considered time span (see Section~\ref{sec:chanmodel}).

The channel estimation, diversity combining, demapping, and decoding are briefly described in the following.

\SubSubSection{Channel Estimation}
\label{sec:chanestalgo}
\vspace{-3mm}

We implement a least-squares error (LSE) channel estimator for the time-domain channel impulse response (CIR) using the $P$ pilot OFDM symbols for each frequency band transmitted in the packet header. For a more general treatment, we let $P$ be a design parameter, but we note that $P=2$ is proposed in \cite{MBOFDM}. The responses in different frequency bands can be estimated separately, since pilot symbols are transmitted for each band. The LSE estimator is chosen instead of minimum-mean-square error (MMSE) estimation because it does not require assumptions regarding the statistical structure of the channel correlations. Furthermore, it has been shown that LSE and MMSE estimation perform almost equally well for cases of practical interest \cite{Cai+Giannakis04}. 

The LSE estimator exploits the fact that the CIR has a maximum of $L\le N$ taps. Starting from (\ref{eq:ofdmsys}), the frequency-domain vector channel estimate can be represented as (cf.\ e.g.\  \cite{Cai+Giannakis04})
\vspace{-3mm}
\begin{equation}
\label{e:chestimate}
        \hat{\ve{H}}=\ve{H}+\ve{E}\;,
\end{equation}
\vspace{-3mm} where the channel estimation error vector ($\cdot^\dagger$ denotes Hermitian transpose)
\begin{equation}
\vspace*{-1mm}
\label{e:esterror}
\ve{E}=\ve{F}_{N\times L} \ve{F}^\dagger_{N\times L}\cdot \frac{1}{P}\sum\limits_{k=1}^P\ve{X}_d^\dagger[k]\ve{N}[k]
\end{equation}
is independent of $\ve{H}$ and zero-mean Gaussian distributed with correlation matrix 
\begin{eqnarray}
        \ve{R}_{\ve{E}\ve{E}} &=& \ve{F}_{N\times L} \ve{F}^\dagger_{N\times L}\left(\frac{\sigma_N^2}{P^2}\sum\limits_{k=1}^P\ve{X}_d^\dagger[k] \ve{X}_d[k]\right)\ve{F}_{N\times L} \ve{F}^\dagger_{N\times L}\nonumber\\
\label{e:corrmatrix}
        &=& \ve{F}_{N\times L}\ve{F}^\dagger_{N\times L}\sigma_N^2/P
\end{eqnarray}
In (\ref{e:esterror}) and (\ref{e:corrmatrix}), $\ve{F}_{N\times L}$ denotes the normalized $N\times L$ FFT matrix with elements $\e^{-\j \mu\nu2\pi/N}/\sqrt{N}$ in row $\mu$ and column $\nu$, and $\sigma_N^2$ is the AWGN variance. For the last step in (\ref{e:corrmatrix}) we assumed the use of constant modulus pilot symbols $|X_i[k]|=1$ (we note that in cases where bit-loading is applied, constant modulus symbols will still be used for the pilots in the packet header). We observe from (\ref{e:chestimate}) and (\ref{e:corrmatrix}) that the LSE channel estimate is disturbed by correlated Gaussian noise with variance
\begin{equation}
\label{e:sigmae}
        \sigma_E^2 = \frac{L}{NP}\sigma_N^2=\eta\sigma_N^2\;.
\end{equation} 
In order to keep complexity low we do not attempt to exploit the correlation, and we further assume that because of interleaving the effect of correlation is negligible. We will refer to parameter $\eta = L/(NP)$ in (\ref{e:sigmae}) when evaluating the performance of Multiband OFDM with imperfect CSI in Sections~\ref{sec:numresimpcsi} and \ref{sec:sim_noloading}. In the remainder of this paper, we assume the maximum impulse response length (i.e., the length of the cyclic prefix) of $L=32$.

\SubSubSection{Diversity Combining, Demapping, and Decoding}
\vspace{-3mm}

Maximum-ratio combining (MRC) \cite{Proakis01} in the case of time and/or frequency spreading (see Section~\ref{sec:mbofdm} and \cite{MBOFDM}) and demapping in the standard BICM fashion \cite{CaireEtAl98} are performed based on the channel estimator output $\hat{\ve{H}}$. The resulting ``soft'' bit metrics are deinterleaved and depunctured. 

Convolutionally coded schemes use a soft-input Viterbi decoder to restore the original bit stream, requiring a decoding complexity of 64 trellis states searched per information bit. Turbo-coded schemes are decoded with 10 iterations of a conventional Turbo decoder using the log-domain BCJR algorithm \cite{Hagenauer96}, with a complexity of roughly $10{\cdot}2{\cdot}2{\cdot}8=320$ trellis states searched per information bit (i.e. 10 iterations of two 8-state component codes, and assuming that the BCJR algorithm is roughly twice as complex as the Viterbi algorithm due to the forward-backward recursion). RA decoding is performed by a turbo-like iterative decoder, using a maximum of 60 iterations and an early-exit criterion which, at relevant values of SNR, reduces the average number of decoder iterations to less than ten \cite{MacKay}. We note that the per-iteration decoding complexity of the RA code is less than that of the Turbo code (since only a 2-state accumulator and a repetition code are used), making the total RA decoder complexity slightly more than the convolutional code but less than the Turbo code. The increased decoder complexities of the Turbo and RA codes, compared to the convolutional code, are reasonable considering the performance gains they provide (see Section \ref{sec:simresults}).

%%%%%%%%%%%%%%%%%%%%%%%%%%%%%%%%%%%%%%%%%%%%%%%%%%%%%%%%%%%%%%%%%%%%%%%%%%%%%%%
\Section{UWB Channel and Diversity Analysis for Multiband OFDM}
\label{sec:Channel}

The UWB channel model developed under IEEE 802.15 \cite{MolischEtAl03} (see Section~\ref{sec:chanmodel}) is a stochastic time-domain model. In this section, we consider a stochastic frequency-domain description, i.e., we include transmitter IFFT and receiver FFT into the channel definition and consider realizations of $\ve{H}$ in (\ref{eq:ofdmsys}). In doing so, we intend to (a) extract the channel parameters relevant for the performance of OFDM-based UWB systems, (b) examine whether the Multiband OFDM proposal is adequate to exploit the channel characteristics, (c) quantify the impact of the different UWB channel types on system performance, and (d) possibly enable a classification of the UWB channel model into more standard channel models used in communication theory.

From (\ref{eq:ofdmsys}) we observe that the OFDM transmit signal experiences a frequency non-selective fading channel with fading along the frequency axis. Thereby, the outer lognormal shadowing term is irrelevant for the fading characteristics as it  affects all tones equally. Hence, the lognormal shadowing term is omitted in the following considerations. Denoting the lognormal term by $G$, we obtain the corresponding \textsl{normalized} frequency-domain fading coefficients as 
\begin{equation}
\label{eq:normchannel}
        \H_i = H_i/G\;.
\end{equation}

\vspace{-5mm}
\SubSection{Marginal Distribution}
\label{sec:marginaldist}
\vspace{-3mm}

The first parameter of interest is the marginal distribution of $\H_i$, i.e., the probability density function (pdf) $p(\H_i)$. 

First, we note that the frequency-domain coefficient $\H_i$ is a zero mean random variable since the time-domain multipath components are zero mean quantities. Furthermore, we have observed that $\H_i$ is, in good approximation, circularly symmetric complex Gaussian distributed, which is explained by the fact that $\H_i$ results from the superposition of many time-domain multipath components. Since these multipath components are mutually statistically independent, the variance of $\H_i$ is independent of the tone index $i$.

Figure~\ref{fig:chanmags} shows measurements of the pdfs $p(|\H_i|)$ of the magnitude frequency-domain gain $|\H_i|$ for the different channel models CM1-CM4, obtained from 10000 independent realizations of channel model. As can be seen, the experimental distributions agree well with the exact Rayleigh distribution of equal variance, which is in accordance with the statements above. We note that similar conclusions regarding the frequency-domain gains were obtained independently in \cite{Wessman04}.
\vspace{-5mm}

\SubSection{Correlation}
\label{sec:correlation}
\vspace{-3mm}

The findings in the previous section indicate that the OFDM signal effectively experiences a (classical) frequency non-selective Rayleigh fading channel (along the OFDM subcarriers). Therefore, knowledge of the second-order channel statistics, i.e., the correlation between different fading coefficients $\H_i$ and $\H_j$, $i\neq j$, is important for the design and assessment of diversity techniques such as coding, interleaving, and frequency hopping, which are envisioned in the Multiband OFDM system. Since coding is performed over all bands, we consider all 3 bands jointly.

\begin{figure}[tb]
\centerline{\input{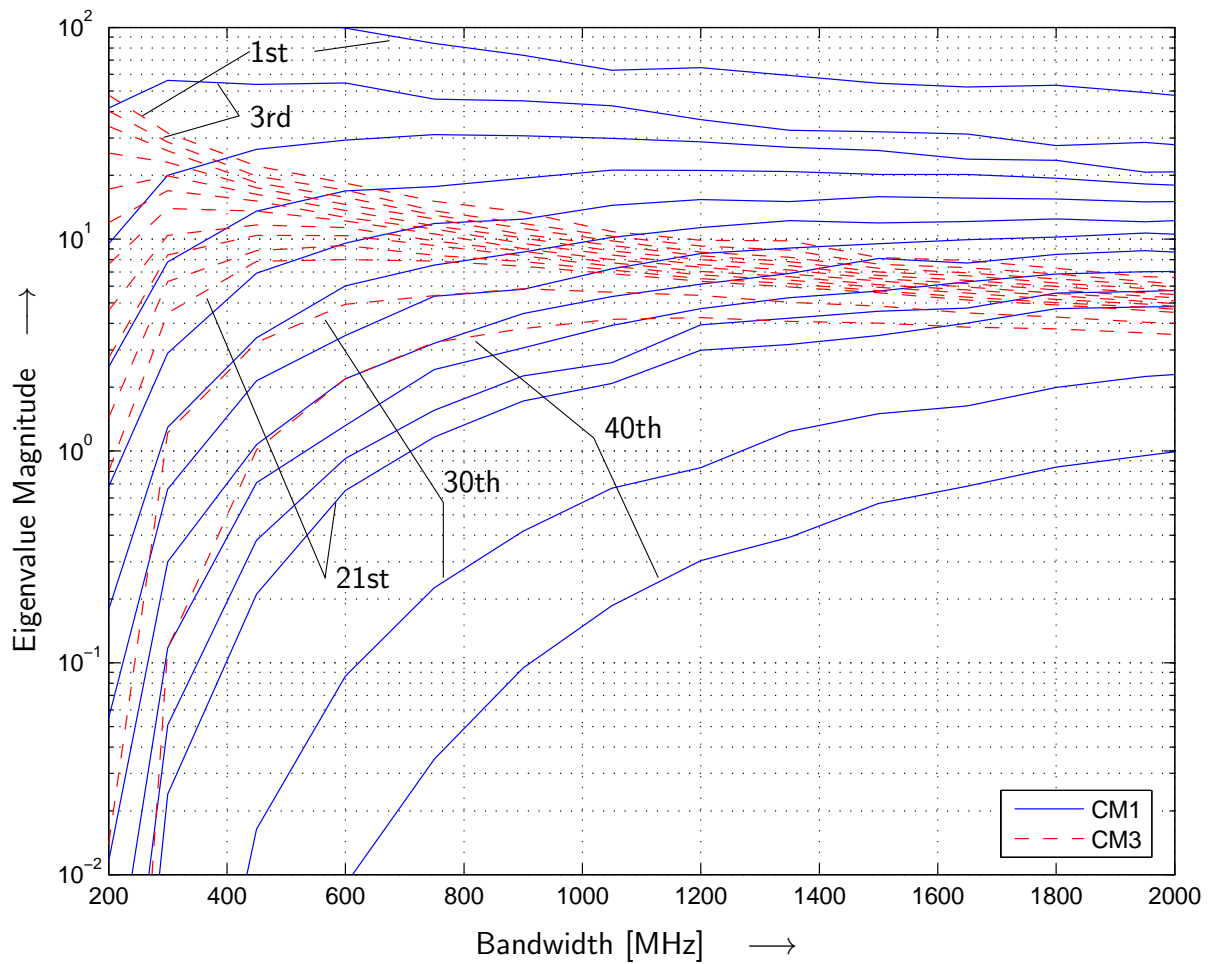}}
\caption{\sf\label{fig:corrmat_eigs_every10th} First 40 ordered eigenvalues of the correlation matrix $\ve{R}_{\vH\vH}$ (every second from 1st to 21st, and the 30th and 40th).}
\end{figure}

As an appropriate figure of merit we examine the ordered eigenvalues of the $3{\cdot}N\times 3{\cdot}N$ correlation matrix $\ve{R}_{\vH\vH}$ of $\vH = [\H_1\ldots \H_{3{\cdot}N}]^T$. Figure~\ref{fig:corrmat_eigs_every10th} shows the first 40 ordered eigenvalues (every second from 1st to 21st, and the 30th and 40th) of the measured $\ve{R}_{\vH\vH}$, which has been obtained from averaging over 1000 channel realizations, as a function of the total employed signal bandwidth. We only show results for channel models CM1 and CM3, which constitute the two extreme cases as the corresponding impulse responses have the least (CM1) and most (CM3) independent multipath components. The respective curve for model CM2 lies in between those for CM1 and CM3.  

From Figure \ref{fig:corrmat_eigs_every10th} we infer the following conclusions: 
\vspace{-4mm}
\begin{enumerate}
\item By increasing the bandwidth of the OFDM signal the diversity order of the equivalent frequency-domain channel, i.e., the number of the significant non-zero eigenvalues of $\ve{R}_{\vH\vH}$, is improved, since, generally, more time-domain multipath components are resolved. However, a 1500~MHz total bandwidth provides already $\ge 40$ (CM3) and $\ge 30$ (CM1) strong diversity branches. This indicates that the 528 MHz bandwidth and 3-band frequency hopping of Multiband OFDM is a favorable compromise between complexity and available diversity.% for CM1.
\vspace{-4mm}
\item Since the system, comprising the convolutional code (see Section \ref{sec:mbofdm}) with free distance $\le 15$ (depending on the puncturing) and spreading factor 1, 2, and 4, can at best exploit diversities of order 15, 30 and 60, respectively, bandwidths of more than 500~MHz per band would only be beneficial for the lowest data-rate modes, and then only for very low error rates. Similar considerations apply to concatenated codes (e.g. Turbo and RA codes as considered in this paper), which do not exceed convolutional codes with spreading in terms of free distance.
\vspace{-4mm}
\item Though CM3 provides higher diversity order than CM1, the latter appears advantageous for high data-rate modes with code puncturing due to its larger first ordered eigenvalues.
\end{enumerate}
\vspace{-3mm}
In summary, we conclude that, given a particular realization of the lognormal shadowing term, the equivalent frequency-domain channel $\ve{H}$ in (\ref{eq:ofdmsys}) is well approximated by a Rayleigh fading channel with relatively high ``fading rate'', which increases from CM1 to CM3.
\vspace{-5mm}

%%%%%%%%%%%%%%%%%%%%%%%%%%%%%%%%%%%%%%%%%%%%%%%%%%%%%%%%%%%%%%%%%%%%%%%%%%%%%%%
\Section{Capacity and Cutoff Rate Analysis}
\label{sec:capeval}

The purpose of this section is to quantify potential data rates and power efficiencies of OFDM-based UWB transmission. Of particular interest here are (a) the channel capacity and cutoff rate,\footnote{\sf It is important to note that the capacity and cutoff rate discussed here are \textsl{constellation-constrained}, i.e., they are calculated assuming a given input constellation with uniform input probabilities.} which are widely accepted performance measures for coded transmission using powerful concatenated codes and convolutional codes, respectively, (b) the influence of the particular channel model (CM1-CM3), and (c) the effect of imperfect channel estimation on these measures. Since coding and interleaving are limited to single realizations of lognormal shadowing, we focus on the notion of \textsl{outage probability}, i.e., the probability that the instantaneous capacity and cutoff rate for a given channel realization $\ve{H}$ fall below a certain threshold. These theoretical performance measures will be compared with simulation results for the Multiband OFDM system in Section \ref{sec:simresults}.

In Section \ref{sec:perfectCSI}, we review the instantaneous capacity and cutoff rate expressions for BICM-OFDM, and extend these expressions to include systems with bit-loading. The required conditional pdf of the channel output is given in Section \ref{sec:noperfectCSI}. Sections \ref{sec:numcapcut} and \ref{sec:loadingcapcut} contain numerical results for systems without and with loading, respectively.
\vspace{-5mm}

\SubSection{Capacity and Cutoff Rate Expressions}
\label{sec:perfectCSI}
\vspace{-3mm}

\SubSubSection{Without Bit-Loading}
\label{sec:capexpr_noloading}
\vspace{-3mm}

The instantaneous capacity in bits per complex dimension of an $N$ tone BICM-OFDM system in a frequency-selective quasi-static channel is given in \cite{EkbalSongCioffi03} (by extending the results of \cite{CaireEtAl98}) as 
\begin{equation}
C(\ve{H}) = m - \frac{1}{N}\sum_{\ell=1}^{m}\sum_{i=1}^{N}\E_{b,Y_i}\left\{\log_2\left(%
        \!\frac{\sum\limits_{X_i\in{\cal X}}\!\!\!p(Y_i|\hat{H}_i,X_i)}%
             {\sum\limits_{X_i\in{\cal X}_b^\ell}\!\!\!p(Y_i|\hat{H}_i,X_i)}%
        \!\right)\right\}.
     \label{eq:cap}
\end{equation}
In (\ref{eq:cap}),  $m$ is the number of bits per symbol, ${\cal X}$ is the signal constellation and ${\cal X}_b^\ell$ is the set of all constellation points $X\in{\cal X}$ whose label has the value $b\in\{0,1\}$ in position $\ell$, $p(Y_i|\hat{H}_i,X_i)$ is the pdf of the channel output $Y_i$ for given input $X_i$ and channel estimate $\hat{H}_i$, and $\E_z\{\cdot\}$ denotes expectation with respect to $z$. For Multiband OFDM, ${\cal X}$ is the QPSK signal constellation and $m=2$ is valid. 

Similarly, we can express the instantaneous cutoff rate in bits per complex dimension as (cf.\ e.g.\ \cite{CaireEtAl98,EkbalSongCioffi03})
\begin{equation}
\label{eq:cut}
R_0(\ve{H}) = m(1 - \log_2(B(\ve{H}) + 1))
\end{equation}
with the instantaneous Bhattacharya parameter ($\bar{b}$ denotes the complement of $b$)
\begin{equation}
B(\ve{H}) = \frac{1}{mN}\sum_{\ell=1}^{m}\sum_{i=1}^{N}%
        \E_{b,Y_i}\left\{\!\sqrt{%
        \frac{\sum\limits_{X_i{\in}{\cal X}_{\bar{b}}^\ell}p(Y_i|\hat{H}_i,X_i)}%
                     {\sum\limits_{X_i{\in}{\cal X}_b^\ell}p(Y_i|\hat{H}_i, X_i)}%
        }\right\}.
\end{equation}

\SubSubSection{With Bit-Loading}
\vspace{-3mm}

The instantaneous capacity in bits per complex dimension of an $N$ tone BICM-OFDM system using loading can be found by extending Eqs. (\ref{eq:cap}) and (\ref{eq:cut}) (following the methodology of \cite{CaireEtAl98,EkbalSongCioffi03}) as
\begin{equation}
C(\ve{H}) = \bar{m} - \frac{1}{N}\sum_{i=1}^{N}\sum_{\ell=1}^{m_i}\E_{b,Y_i}\left\{\log_2\left(%
        \frac{\sum\limits_{X_i\in{\cal X}_i}p(Y_i|\hat{H}_i,X_i)}%
             {\sum\limits_{X_i\in{\cal X}_{i,b}^\ell}p(Y_i|\hat{H}_i,X_i)}%
        \right)\right\}.
     \label{eq:loadingcap}
\end{equation}
In (\ref{eq:loadingcap}),  $\bar{m}$ is the average number of bits/symbol ($\bar{m}=2$ throughout this paper), $m_i$ and ${\cal X}_i$ are the number of bits per symbol and the signal constellation for the $i^{\mathrm{th}}$ tone, respectively, and ${\cal X}_{i,b}^\ell$ is the set of all constellation points $X\in{\cal X}_i$ whose label has the value $b\in\{0,1\}$ in position $\ell$. 

Similarly, we can express the instantaneous cutoff rate for bit-loading systems in bits per complex dimension as 
\begin{equation}
\label{eq:loadingcut}
R_0(\ve{H}) = \bar{m}(1 - \log_2(B(\ve{H}) + 1))
\end{equation}
with the instantaneous Bhattacharya parameter
\begin{equation}
B(\ve{H}) = \frac{1}{N}\sum_{i=1}^{N}\sum_{\ell=1}^{m_i}\frac{1}{m_i}%
        \E_{b,Y_i}\left\{\sqrt{%
        \frac{\sum\limits_{X_i{\in}{\cal X}_{i,\bar{b}}^\ell}p(Y_i|\hat{H}_i,X_i)}%
                     {\sum\limits_{X_i{\in}{\cal X}_{i,b}^\ell}p(Y_i|\hat{H}_i, X_i)}%
        }\right\}.
\end{equation}

\SubSection{Conditional PDF}
\label{sec:noperfectCSI}
\vspace{-3mm}

In order to calculate capacity and cutoff rate, we require the conditional pdf $p(Y_i|\hat{H}_i,X_i)$. In the case of perfect CSI we have $\hat{H}_i=H_i$, and $p(Y_i|\hat{H}_i,X_i)$ is a Gaussian pdf with mean $H_iX_i$ and variance $\sigma_N^2$.

We now obtain $p(Y_i|\hat{H}_i,X_i)$ for the more realistic case of imperfect CSI assuming the application of LSE channel estimation as described in Section \ref{sec:mbofdmrec}. According to the results of Section \ref{sec:marginaldist} and since channel estimation is performed for one realization $G$ of the lognormal shadowing term, we further assume zero-mean circularly symmetric Gaussian distributed channel coefficients $H_i$ with variance $\sigma_H^2=G^2$ (see Eq.\ (\ref{eq:normchannel})). This means that $\hat{H}_i$ is also zero-mean Gaussian distributed with variance $\sigma_{\hat{H}}^2=\sigma_H^2+\sigma_E^2$ (see Eqs.\ (\ref{e:chestimate}) and (\ref{e:sigmae})). Let $\mu$ be the correlation between $H_i$ and $\hat{H}_i$,  
\begin{equation}
\mu =\frac{\E_{H_i,\hat{H}_i}\{H_i\hat{H}_i^\ast\}}{\sigma_{H}\sigma_{\hat{H}}}=  \sqrt{\frac{\sigma^2_H}{\sigma_E^2 + \sigma_H^2}} = \sqrt{\frac{\gamma}{\gamma + \eta}}\;,
\end{equation}
where $(\cdot)^\ast$ denotes the complex conjugate, $\eta$ is defined in (\ref{e:sigmae}), and ${\gamma} = \sigma_H^2/\sigma_N^2$ is the SNR. Then, we can arrive via algebraic manipulations at (cf.\ e.g.\ \cite{WuEtAl04})
\begin{equation}
\label{eq:condpdf_impcsi}
p(Y_i|\hat{H}_i,X_i) = \frac{1}{{\pi}(\sigma_N^2(\eta\mu^2+1))}\exp\left(-\frac{{|Y_i-X_i \hat{H}_i\mu^2|}^2}{\sigma_N^2(\eta\mu^2+1)}\right)\;.
\end{equation}

The Gaussian density of (\ref{eq:condpdf_impcsi}) implies that the system with imperfect CSI can be seen as a system with perfect CSI at an equivalent SNR of
\begin{equation}
\label{eq:equivSNR}
\gamma_e = \frac{\E_{\hat{H}_i}\{|\hat{H}_i|^2\}\mu^4}{\sigma_N^2(\eta\mu^2 + 1)}= \frac{{\gamma}}{\eta\left(1 + \frac{1}{{\gamma}}\right) + 1}\;.
\end{equation}
We note that in the high SNR regime the loss due to estimation error reaches a constant value of $1/(\eta+1)$.

\SubSection{Numerical Results --- No Loading}
\label{sec:numcapcut}
\vspace{-3mm}

We evaluated expressions (\ref{eq:cap}) and (\ref{eq:cut}) via Monte Carlo simulation using 1000 realizations of each UWB channel model CM1-CM3. To keep the figures legible, we present representative results for CM1 and CM3 only. The performance of CM2 (not shown) is between that of CM1 and CM3 (cf.\ also Section \ref{sec:correlation}). For comparison we also include results for independent and identically distributed (i.i.d.) Rayleigh fading on each tone and an outer lognormal shadowing term identical to that of the UWB models (labeled as ``Rayleigh + LN'').

\SubSubSection{Perfect CSI}
\vspace{-3mm}

\begin{figure}[tbp]
\centerline{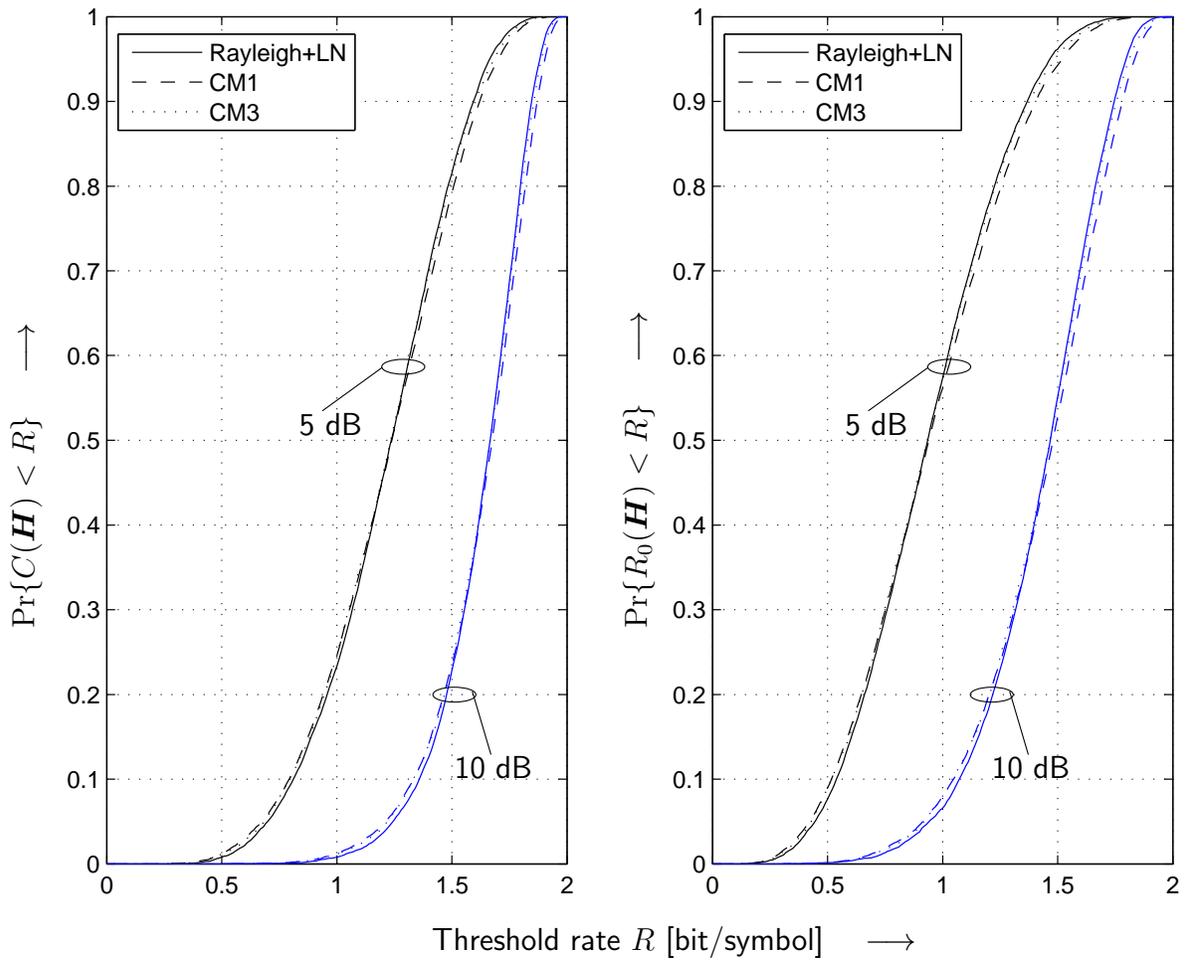}
\caption{\sf\label{fig:cap_ccdf} %
Outage probability for $10\log_{10}(\Es/{\cal N}_0)=5$~dB and 10~dB and perfect CSI. Left: Outage capacity. Right: Outage cutoff rate.}
\end{figure}

First, we consider the case of perfect CSI. Figure~\ref{fig:cap_ccdf} shows the outage capacity $\Pr\{C(\ve{H})<R\}$ (left) and cutoff rate $\Pr\{R_0(\ve{H})<R\}$ (right) as a function of the threshold rate $R$ for $10\log_{10}(\Es/{\cal N}_0)=5$~dB and 10~dB, respectively, where $\Es$ is the average received energy per symbol and ${\cal N}_0$ denotes the two-sided power spectral density of the complex noise. 

It can be seen that both capacity and cutoff rate for the UWB channel models are similar to the respective parameters of an i.i.d.\ Rayleigh fading channel with additional lognormal shadowing. In fact, the curves for CM3, which provides the highest diversity (see Section \ref{sec:correlation}), are essentially identical to those for the idealized i.i.d.\ model. The high diversity provided by the UWB channel also results in relatively steep outage curves, which means that transmission reliability can be considerably improved by deliberately introducing coding redundancy. This effect is slightly more pronounced for the capacity measure relevant for more powerful coding. On the other hand, the effect of shadowing, which cannot be averaged out by coding, causes a flattening towards low outage probabilities $\le 0.1$. In the high outage probability range we note that CM1 is slightly superior to CM3, which is due to the large dominant eigenvalues of CM1 identified in Section \ref{sec:correlation}. 

\begin{figure}[tbp]
\centerline{\input{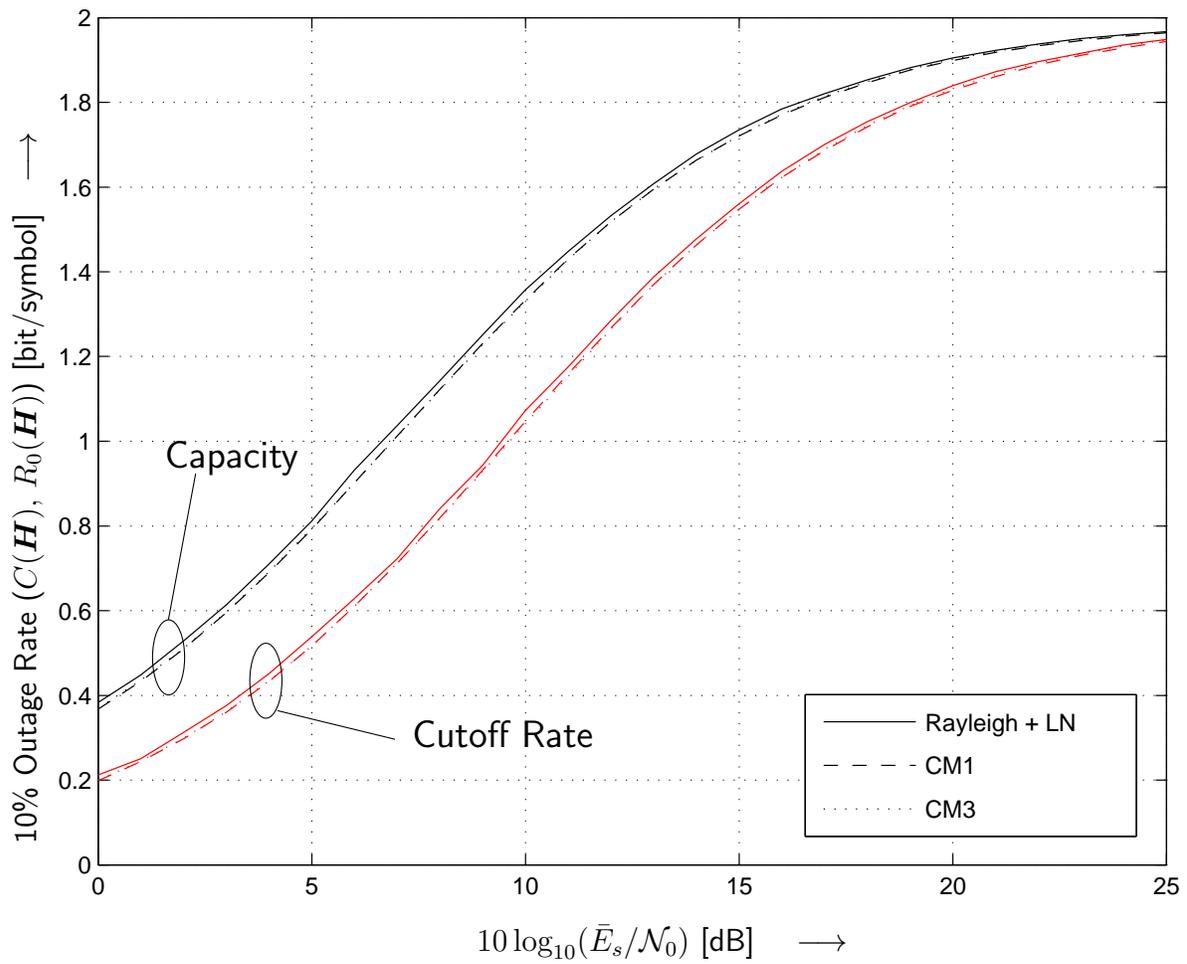}}
\caption{\sf\label{fig:cap_outages} %
10\% outage capacity and cutoff rate for perfect CSI.}
\end{figure}

In Figure~\ref{fig:cap_outages} we consider the 10\% outage\footnote{\sf We note that 10\% outage is a typically chosen value for UWB systems and the considered channel model \cite{MBOFDM}.} capacity and cutoff rate as a function of the SNR\\
$10\log_{10}(\Es/{\cal N}_0)$.  Again we note the close similarity between the UWB channel models and the i.i.d.\ Rayleigh fading channel with lognormal shadowing. A comparison of the capacity with the corresponding cutoff rate curves indicates that decent gains of 2.5~dB to 3~dB in power efficiency can be anticipated by the application of more powerful capacity approaching codes such as Turbo or RA codes (see also the simulation results in Section \ref{sec:sim_noloading}) instead of the convolutional codes proposed in \cite{MBOFDM} which usually perform in the vicinity of the cutoff rate.

\SubSubSection{Imperfect CSI}
\label{sec:numresimpcsi}
\vspace{-3mm}

\begin{figure}[tbp]
\centerline{\input{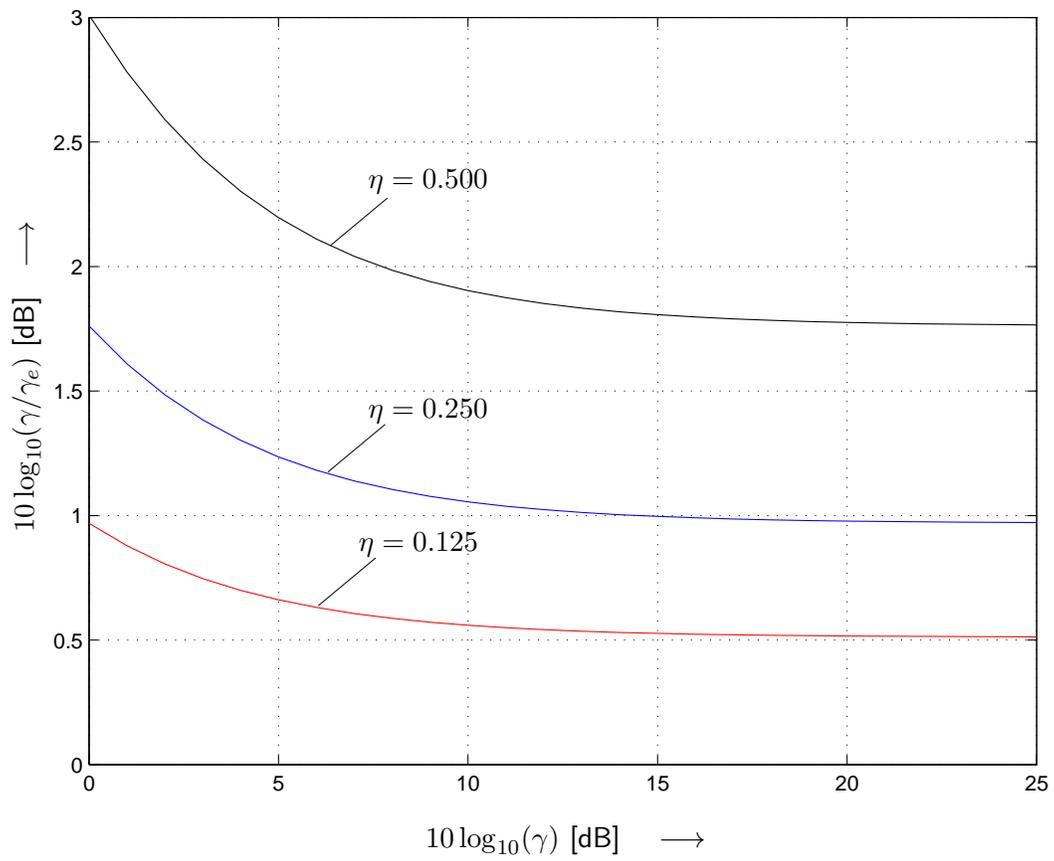}}
\caption{\sf\label{fig:cut_outages_impCSI} %
Loss in SNR due to LSE channel estimation with different $\eta$ according to (\protect\ref{eq:equivSNR}).}
\end{figure}

Figure \ref{fig:cut_outages_impCSI} shows the SNR loss due to LSE channel estimation according to Eq.\ (\ref{eq:equivSNR}) with various values of $\eta$. For reference, the Multiband OFDM system uses $P = 2$, $N = 128$, and so choosing $L = 32$ (equal to the cyclic prefix length) leads to $\eta=0.125$. 

We can see from  Figure \ref{fig:cut_outages_impCSI} that the performance penalty $10\log_{10}(\gamma/\gamma_e)$ due to imperfect CSI is about 0.5~dB in the range of interest for the Multiband OFDM system. The actual loss in $\Es/{\cal N}_0$ is slightly different, since $\gamma$ in Figure \ref{fig:cut_outages_impCSI} is for a fixed lognormal shadowing and the actual $\Es/{\cal N}_0$ loss must be obtained by averaging over the lognormal pdf. However, we can see from Figure \ref{fig:cut_outages_impCSI} that the SNR loss is relatively constant for relevant values of $\gamma$, which (since the lognormal shadowing has a 0~dB mean), results in $\Es/{\cal N}_0$ loss of approximately $10\log_{10}(\gamma/\gamma_e)$. Reducing the channel estimation overhead to $P=1$ ($\eta=0.25$) could be an interesting alternative for short packets, as the additional loss is only about 0.5~dB (in terms of required energy per information bit $\bar{E}_b$ the loss is even smaller). Further reduction of pilot tones is not advisable as the gains in throughput are outweighed by the losses in power efficiency.

\SubSection{Numerical Results with Bit-Loading}
\label{sec:loadingcapcut}
\vspace{-3mm}

In this section, we examine the capacity and cutoff rate of systems employing the Piazzo and CCB loading algorithms. We evaluated expressions (\ref{eq:loadingcap}) and (\ref{eq:loadingcut}) via Monte Carlo simulation as discussed in Section~\ref{sec:numcapcut}.

\SubSubSection{No Clustering}
\label{sec:noclustercapcutres}
\vspace{-3mm}

\begin{figure}[tbp]
\centerline{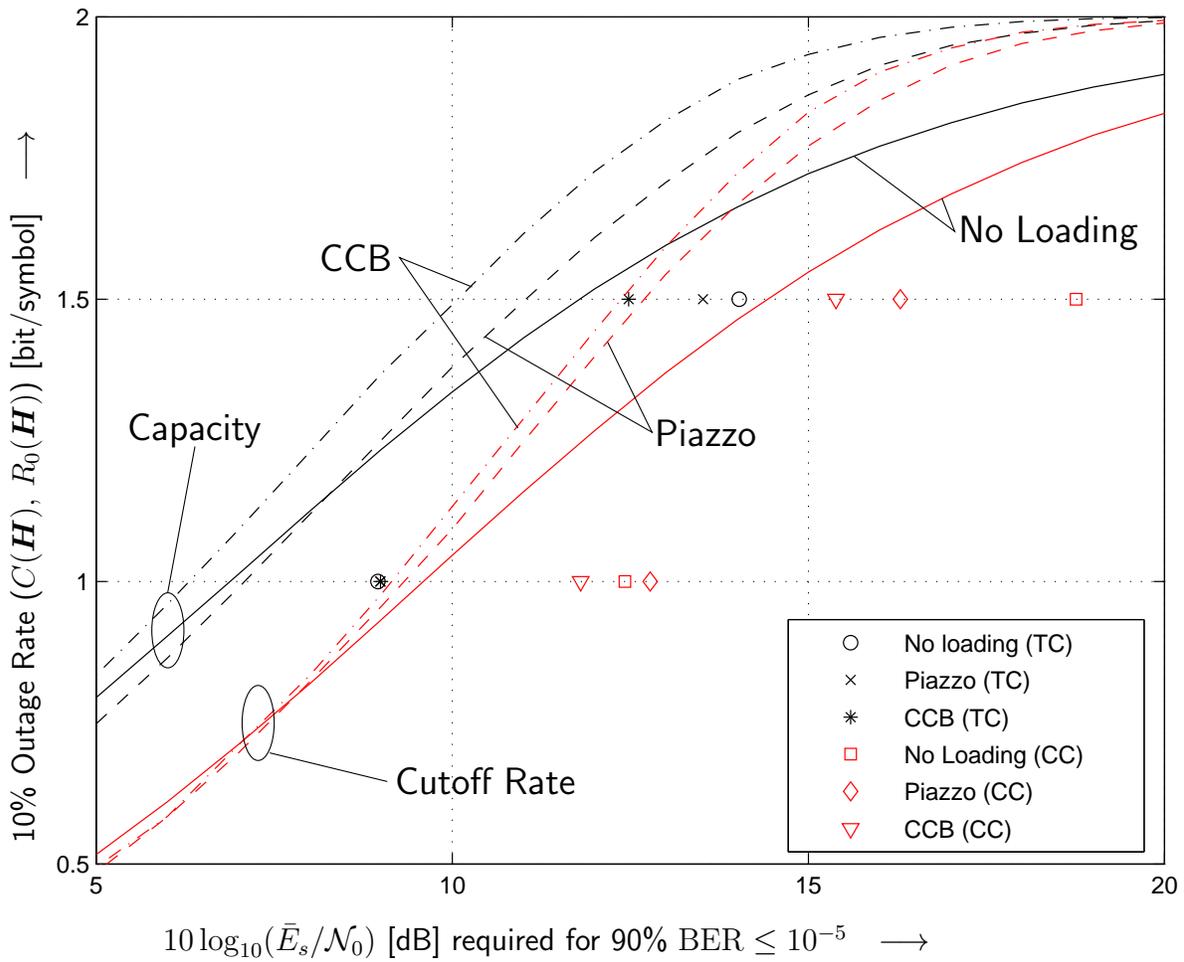}
\caption{\sf\label{fig:loading} %
10\% outage capacity and cutoff rate with and without loading for CM1 (lines). $10\log_{10}(\Es/{\cal N}_0)$ required to achieve $\ber\le 10^{-5}$ for the 90\% best channel realizations using convolutional and Turbo codes, with and without loading (markers).}
\end{figure}

Figure~\ref{fig:loading} (lines) shows the 10\% outage capacity and cutoff rates for the CM1 channel using the Piazzo and CCB loading algorithms. (The markers in this figure will be discussed in Section~\ref{sec:sim_loading}). It should be noted that $\Es$ is not adjusted to account for tones carrying 0 bits, because we assume operation at FCC transmit power limits, precluding the re-allocation of power from unused tones to other subcarriers (which would put the transmit power spectral density beyond the allowed limits). We also do not adjust for the overhead associated with the feedback of loading information from the receiver to the transmitter. For high rates, both the CCB and the Piazzo loading algorithms provide a gain of several dB in capacity and in cutoff rate compared to the unloaded case, and this gain grows with increasing rate and $\Es/{\cal N}_0$. The Piazzo algorithm is sub-optimal because it considers only the relative SNR between tones, and loads according to BER using a power minimization criterion. This loading strategy is not guaranteed to produce an increased channel capacity (or cutoff rate). On the other hand, the CCB algorithm requires knowledge of the actual SNR values of each tone and loads according to their approximate capacities, resulting in an increased channel capacity for all SNR values and an improved performance compared to Piazzo loading.

\SubSubSection{Clustering}
\label{sec:clustercapcutres}
\vspace{-3mm}

\begin{figure}[tbp]
\centerline{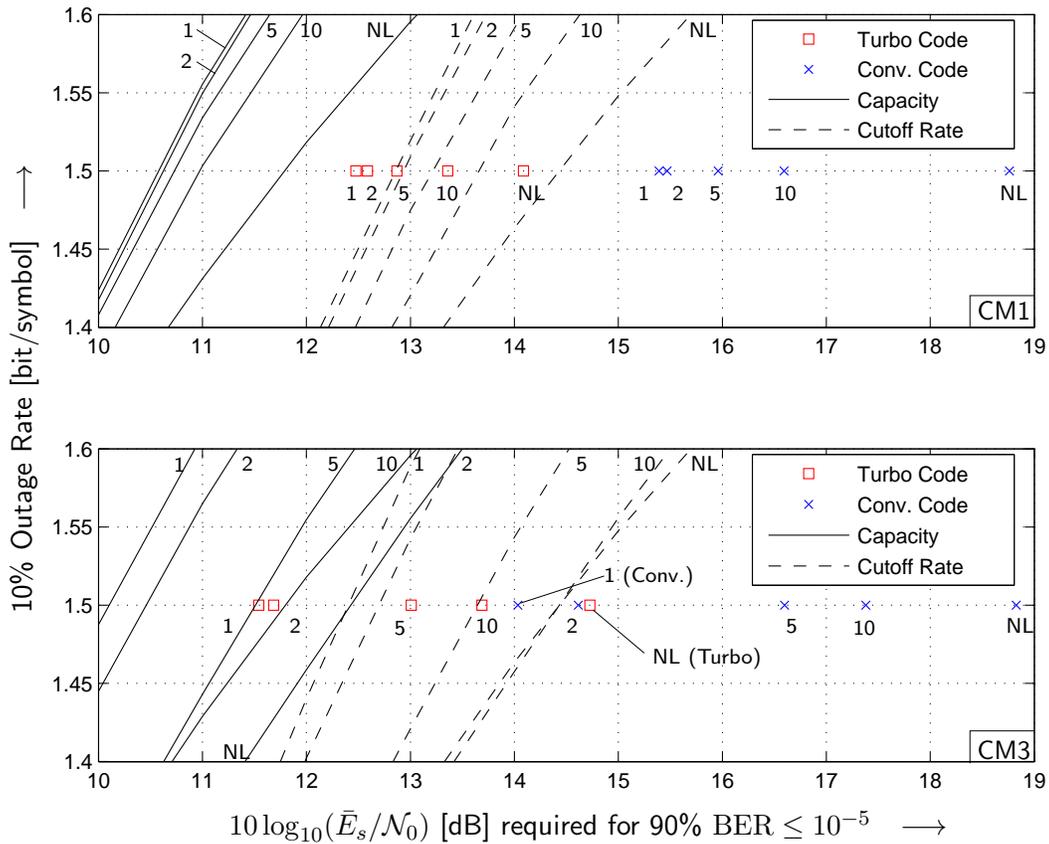}
\caption{\sf\label{fig:clusterloading} %
Lines: 10\% outage capacity (solid) and cutoff rate (dashed) for clustered CCB loading (cluster sizes $D\in\{1,2,5,10\}$) and for non-loaded QPSK (``NL''). Markers: $10\log_{10}(\Es/{\cal N}_0)$ required to achieve $\ber\le 10^{-5}$ for the 90\% best channel realizations using Turbo codes ($\Box$ markers) and convolutional codes (x markers). Channels CM1 (top) and CM3 (bottom).}
\end{figure}

We next consider the application of clustered loading using the modified CCB algorithm as described in Section~\ref{sec:modulation}. Figure~\ref{fig:clusterloading} shows the 10\% outage capacity (solid lines) and cutoff rate (dashed lines) for various values of cluster size $D$, for channels CM1 and CM3. Also included for comparison are the non-clustered loading ($D\!=\!1$) and unloaded (all-QPSK) curves. As the cluster size $D$ increases the attainable rates decrease because the modulation scheme chosen for each cluster is not optimal for all tones in the cluster. This loss is slightly more pronounced for the cutoff rate than for the capacity, which indicates that when using clustered loading we should expect more performance degradation with convolutional codes than with Turbo codes (see also Section~\ref{sec:sim_loading}). The performance degradation with increasing cluster size is higher for CM3 than for CM1, which can be predicted from the correlation matrix results of Section~\ref{sec:correlation}. Specifically, we note from Figure \ref{fig:corrmat_eigs_every10th} that the frequency responses of adjacent subcarriers are more correlated (fewer significant eigenvalues) in CM1 and less correlated (more significant eigenvalues) in CM3. The less correlated the tones of a cluster are, the higher the average mismatch between the optimal modulation for each tone (i.e. that chosen by the non-clustered loading algorithm) and the fixed modulation chosen for the cluster. The higher average mismatch on CM3 results in lower performance when clustered loading is applied. 
\vspace{-5mm}

%%%%%%%%%%%%%%%%%%%%%%%%%%%%%%%%%%%%%%%%%%%%%%%%%%%%%%%%%%%%%%%%%%%%%%%%%%%%%%%
\Section{Simulation Results}
\label{sec:simresults}
\vspace{-3mm}

In Section \ref{sec:sim_noloading}, we study Turbo, RA, and convolutional coding without bit-loading. We examine channel CM1 with four different transmission modes with data rates of 80, 160, 320, and 480~Mbps corresponding to 0.25, 0.50, 1.00, and 1.50 bit/symbol, respectively. In the simulations, detection is performed with perfect CSI as well as with LSE channel estimation using $\eta=0.125$. We then turn to the performance of systems with loading in Section \ref{sec:sim_loading}. Based on the results of the information-theoretic analysis of Section \ref{sec:loadingcapcut}, we restrict our attention to rates $\ge 1.00$ bit/symbol, where we expect loading algorithms to yield performance gains. We concentrate on Turbo and convolutional codes for this section. The simulation results presented in these two sections are the worst-case $10\log_{10}(\Es/{\cal N}_0)$ values required to achieve $\ber{\le}10^{-5}$ for the best 90\% of channel realizations over a set of 100 channels (i.e. they are simulation results corresponding to 10\% outage). 

In Section \ref{sec:range}, we briefly summarize the power efficiency gains and attendant range improvements expected from the application of the system extensions we have proposed.
\vspace{-2mm}

\SubSection{No Loading}
\label{sec:sim_noloading}
\vspace{-3mm}

\begin{figure}[tbp]
\centerline{\input{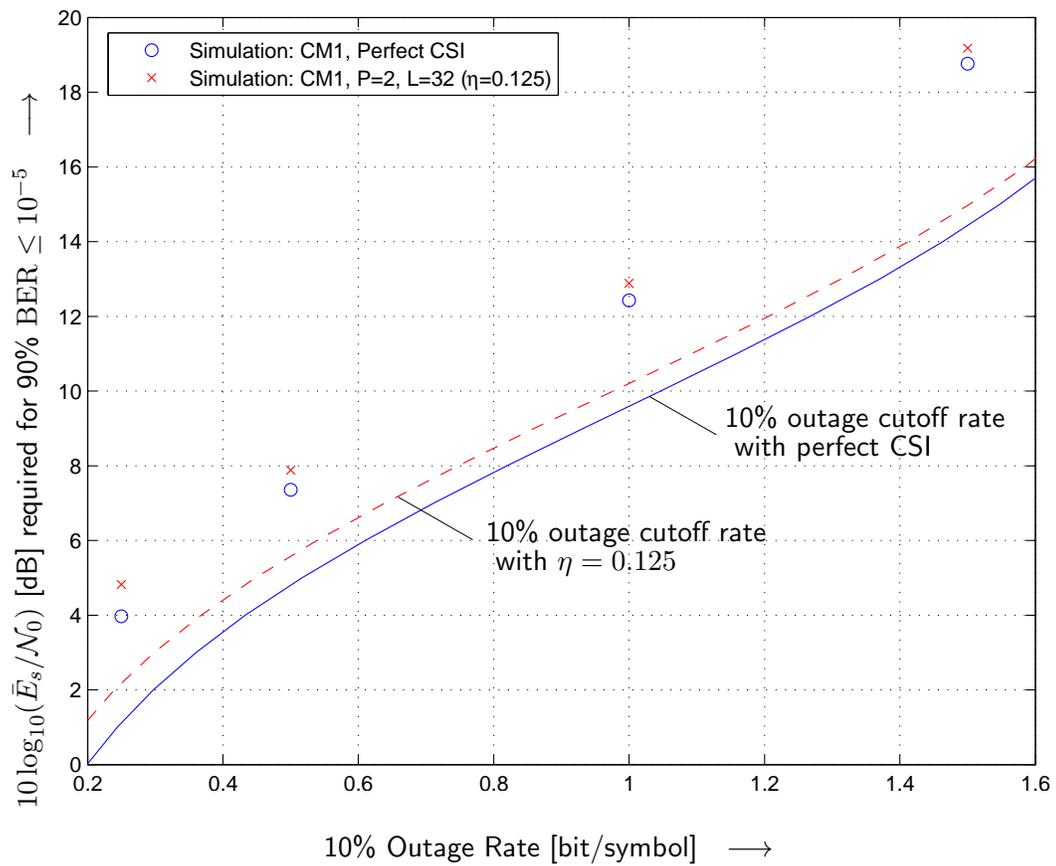}}
\caption{\sf\label{fig:bersim} %
$10\log_{10}(\Es/{\cal N}_0)$ required to achieve $\ber\le 10^{-5}$ for the 90\% best channel realizations using convolutional codes (markers). For comparison: 10\% outage cutoff rate (lines). Channel model CM1 and LSE channel estimation.}
\end{figure}

Figure~\ref{fig:bersim} (markers) shows SNR points when using convolutional codes (as in the Multiband OFDM proposed system), together with the corresponding 10\% outage cutoff rate curves. We observe that the simulated SNR points are approximately 3~dB to 4~dB from the cutoff-rate curves, which is reasonable for the channel model and coding schemes under consideration. These results (a) justify the relevance of the information-theoretic measure and (b) confirm the coding approach used in Multiband OFDM. More specifically, the diversity provided by the UWB channel is effectively exploited by the chosen convolutional coding and interleaving scheme. Furthermore, the system with LSE channel estimation performs within 0.5--0.7~dB of the perfect CSI case as was expected from the cutoff-rate analysis (see also the discussion in Section \ref{sec:numresimpcsi} on the relationship between the loss $10\log_{10}(\gamma/\gamma_e)$ and the $10\log_{10}(\Es/{\cal N}_0)$ loss).

\begin{figure}[tbp]
\centerline{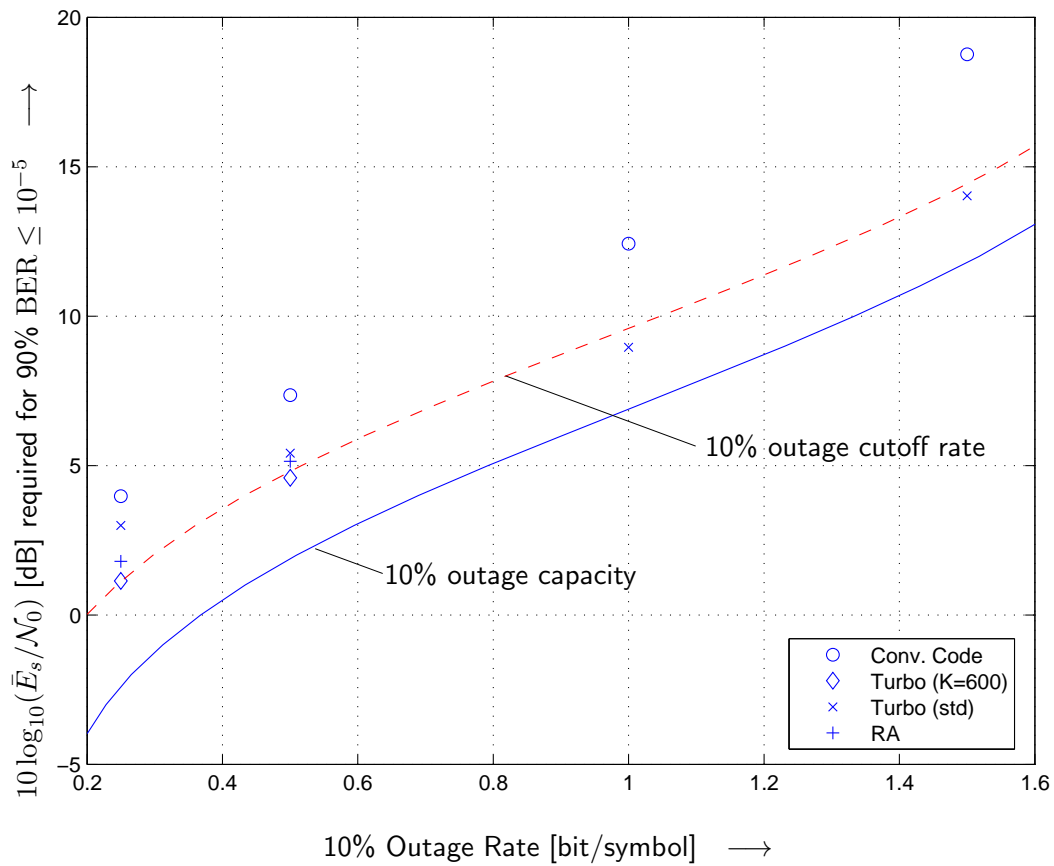}
\caption{\sf\label{fig:BERcap} %
$10\log_{10}(\Es/{\cal N}_0)$ required to achieve $\ber\le 10^{-5}$ for the 90\% best channel realizations using Turbo Codes, RA codes, and convolutional codes (markers). For comparison: 10\% outage capacity and cutoff rate (lines). Channel model CM1 and perfect CSI.}
\end{figure}

We next consider the Turbo and RA coding schemes. Figure~\ref{fig:BERcap} (markers) shows the simulation results for Turbo and RA codes on channel CM1 with perfect CSI, as well as the convolutional code results for comparison. We also show the corresponding 10\% outage capacity and cutoff rate curves. Turbo codes give a performance gain of up to 5~dB over convolutional codes, and perform within 2.5~dB of the channel capacity, depending on the rate. At rates of 0.25 and 0.50 bit/symbol, Turbo code interleaver sizes compatible with the channel interleaver design of the Multiband OFDM proposal (the ``std'' points) incur a performance penalty of 1--2~dB compared with the longer block length (``K=600'') points. Repeat-accumulate codes have a performance roughly 1~dB worse than the long block-length Turbo codes, but the RA codes are both (a) compatible with the Multiband OFDM channel interleaver, and (b) less complex to decode. They are thus a good candidate for low-rate Multiband OFDM transmission. 
%However, the design of good RA codes for high rates is a difficult problem, and thus Turbo codes are best suited for use at higher rates. 

\vspace{-2mm}
\SubSection{With Loading}
\label{sec:sim_loading}
\vspace{-3mm}

Figure~\ref{fig:loading} (markers) shows the simulation results for Turbo codes and for convolutional codes, using both the CCB and Piazzo loading algorithms on channel CM1 with perfect CSI. At 1.00 bit/symbol and using convolutional codes, we see a performance gain of less than 1~dB using CCB loading, and a slight performance degradation using Piazzo loading. Performance using Turbo codes at 1.00 bit/symbol is relatively constant regardless of loading. However, at 1.50 bit/symbol we see gains of approximately 1.5~dB for Turbo codes and almost 4~dB for convolutional codes when CCB loading is used. The gains using the Piazzo algorithm are approximately 1~dB less, as predicted by the capacity analysis of Section \ref{sec:loadingcapcut}. Finally, we note that at 1.50 bit/symbol the system employing CCB loading and Turbo codes is approximately 6~dB better than the unloaded convolutionally coded system, and performs within approximately 2.5~dB of the channel capacity.

In Figure~\ref{fig:clusterloading} (markers) we consider the performance of clustered loading with Turbo codes and with convolutional codes for 1.50 bit/symbol on the CM1 and CM3 channels with perfect CSI. As predicted by information-theoretic analysis, clustered loading incurs a performance penalty with increasing cluster size $D$. We note that Turbo codes suffer a smaller performance degradation (relative to $D\!=\!1$) than convolutional codes, because the more powerful Turbo code is better suited to handle the mismatched modulation (as discussed in Section \ref{sec:clustercapcutres}). The performance degradation is larger for CM3 due to that channel model's lower correlation between adjacent subcarrier frequency responses and resultant larger loading mismatch. However even $D\!=\!10$ loading provides performance gains for both channels and code types. Cluster size $D\!=\!2$ is a good tradeoff point for both Turbo and convolutional codes, allowing for feedback reduction by a factor of 2 with losses of approximately 0.1~dB for CM1 and 0.4~dB for CM3. Cluster sizes as large as $D\!=\!5$ could be used with Turbo codes, depending on the required power efficiency and expected channel conditions.
\vspace{-2mm}

\SubSection{Range Improvements from Turbo Codes and Loading}
\label{sec:range}
\vspace{-3mm}

Table~\ref{table:range} lists the gains in required $10\log_{10}(\Es/{\cal N}_0)$ and percentage range increases on channel CM1 for various combinations of the extensions we have proposed.  We assume a path loss exponent of $d=2$, as in \cite{BatraEtAl04}. We can see that bit loading alone provides up to 47\% increase in range, Turbo codes without loading provide a 71\% increase, and the combination of Turbo codes and loading allows for a 106\% increase in range. Furthermore, the use of clustered loading with $D\!=\!2$ only reduces these range improvements by 1\% to 3\% over the non-clustered case, while providing reduced-rate feedback and lower computational complexity.
\vspace{-4mm}

%%%%%%%%%%%%%%%%%%%%%%%%%%%%%%%%%%%%%%%%%%%%%%%%%%%%%%%%%%%%%%%%%%%%%%%%%%%%%%%
\Section{Conclusions}
\label{sec:conclusions}
\vspace{-3mm}

In this paper, the application of Multiband OFDM for UWB communication has been analyzed. We have shown that the UWB channel model developed under IEEE 802.15 is seen by OFDM systems in the frequency domain as Rayleigh fading with additional shadowing. The 528~MHz signal bandwidth chosen for Multiband OFDM essentially captures the diversity provided by the UWB channel. As a result, we have found that the information-theoretic limits of the UWB channel are similar to those of a perfectly interleaved Rayleigh fading channel with shadowing. The BICM-OFDM scheme proposed for Multiband OFDM performs close to the outage cutoff-rate measure and is thus well suited to exploit the available diversity. The application of stronger coding, such as Turbo codes or Repeat-Accumulate codes, improves power efficiency by up to 5~dB, depending on the data rate. Bit-loading algorithms provide additional performance gains for high data rates, and a simple clustering scheme allows for reduced-rate feedback of loading information depending on the channel conditions and required power efficiency. Finally, a simple LSE channel estimator has been shown to enable performance within 0.5--0.7~dB of the perfect CSI case for the proposed Multiband OFDM system.

%%%%%%%%%%%%%%%%%%%%%%%%%%%%%%%%%%%%%%%%%%%%%%%%%%%%%%%%%%%%%%%%%%%%%%%%%%%%%%%
%%\clearpage
\renewcommand{\baselinestretch}{1.15}\small\makeatletter
\def\@listi{\leftmargin\leftmargini
            \parsep 5\p@  \@plus2.5\p@ \@minus\p@
            \topsep 10\p@ \@plus4\p@   \@minus6\p@
            \itemsep-5\p@  \@plus1\p@ \@minus0\p@}
\makeatother
\bibliography{IEEEabrv,local}
\bibliographystyle{IEEEtran}

%%%%%%%%%%%%%%%%%%%%%%%%%%%%%%%%%%%%%%%%%%%%%%%%%%%%%%%%%%%%%%%%%%%%%%%%%%%%%%%
\newpage

\clearpage
\begin{figure}[tbp]
\centerline{\input{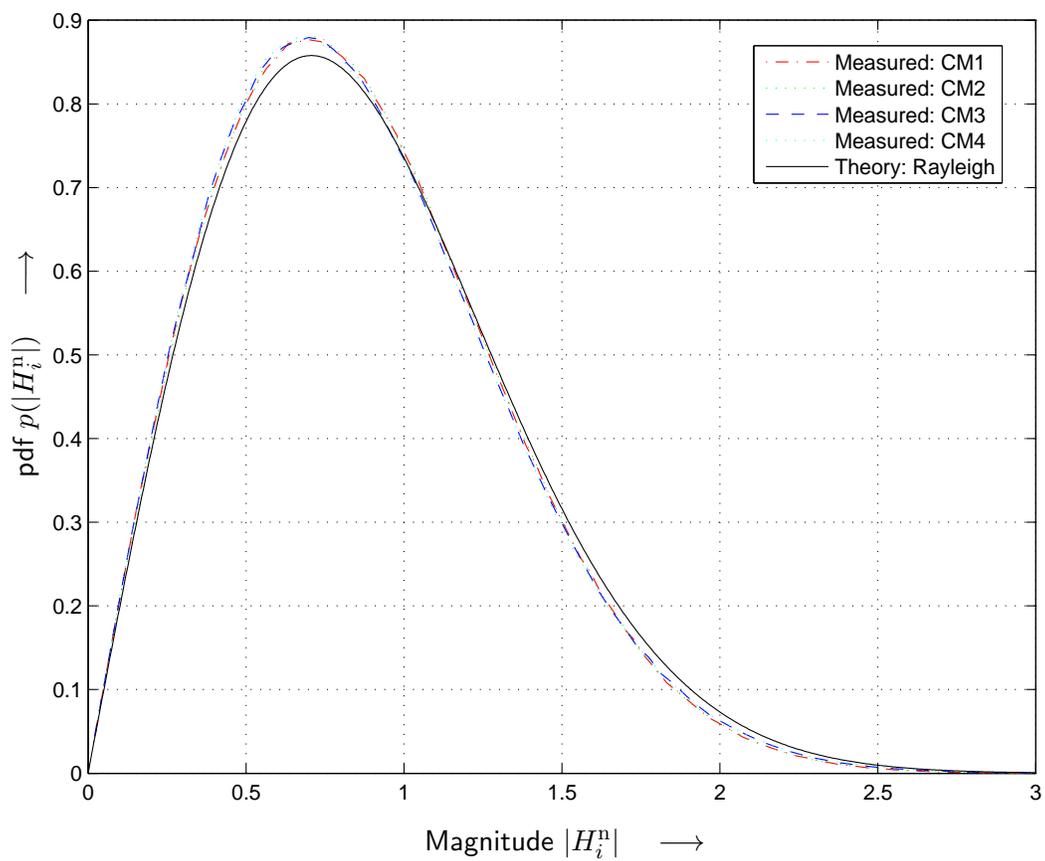}}
  \caption{\sf\label{fig:chanmags}\def\baselinestretch{1.0}
Distributions of normalized channel magnitude $|\H_i|$ for channel types CM1-CM4. For comparison: Rayleigh distribution with same variance.}
\end{figure}

\clearpage

\clearpage

\clearpage

\clearpage

\clearpage

\clearpage

\clearpage

\clearpage

\clearpage
\begin{table}[tbp]
\caption{\sf \label{table:range} Power efficiency gains and range increases available using some of the extensions considered, compared to the Multiband OFDM standard proposal. Channel CM1, rate 1.50 bit/symbol (480 Mbps), path loss exponent $d\!=\!2$. $10\log_{10}(\Es/{\cal N}_0)$ values are those required to achieve $\ber\le 10^{-5}$ for the 90\% best channel realizations. (CC: convolutional code, TC: Turbo code).}
\vspace{5mm}
\centerline{
\begin{tabular}{@{\extracolsep{\fill}}|l|c|c|c|}
\hline
\sf System & $10\log_{10}(\Es/{\cal N}_0)$ & \sf Gain (dB) & \sf \% range increase \\
\hline
\hline
\sf CC, no loading & \sf 18.76 & \sf $-$ & \sf $-$ \\  % final
\sf (Standard Proposal) & & & \\
\hline
\sf CC, CCB loading & \sf 15.38 & \sf 3.38 & \sf 47 \% \\  % final
\hline
\sf CC, $D=2$ clustered loading & \sf 15.47 & \sf 3.29 & \sf 46 \% \\
\hline
\sf TC, no loading & \sf 14.09 & \sf 4.67 & \sf 71 \% \\  % final
\hline
\sf TC, CCB loading & \sf 12.48 & \sf 6.28  & \sf 106 \% \\  % final
\hline
\sf TC, $D=2$ clustered loading & \sf 12.58 & \sf 6.18 & \sf 103 \% \\  % final
\hline
\end{tabular}
}
\end{table}

\end{document}